\documentclass[prd,a4paper,nofootinbib, preprintnumbers, amsmath,multicol,12pt]{revtex4}

\usepackage{color}
\usepackage{latexsym}
\usepackage{epsfig}
\usepackage{amsmath}
\usepackage{amssymb}
\usepackage{graphicx}
\usepackage{bm}
\makeatletter

\newcommand\ee{\end{equation}}
\newcommand\be{\begin{equation}}
\newcommand\eea{\end{eqnarray}}
\newcommand\bea{\begin{eqnarray}}

\def\beq{\begin{equation}}
\def\eeq{\end{equation}}

\begin{document}
\setcounter{page}{0}
\thispagestyle{empty}

\preprint{\footnotesize CP3-Origins-2012-22}
\preprint{\footnotesize DIAS-2012-23}

%\begin{titlepage}

~\vspace{1cm}

\begin{center}

{\Large\bf \color{red}  
Higgs boson and Top quark masses \\ as tests of Electroweak Vacuum Stability }
\vspace{1cm}

{\large  %\sl 
Isabella Masina$^{a,b}$  }
\\[.5cm]
{\normalsize \small { \sl
 $^{a}$ Dipartimento di Fisica dell'Universit\`a di Ferrara and INFN Sezione di Ferrara, \\ \vspace{-.2cm}Via Saragat 1, I-44100 Ferrara, Italy}}\\
{\normalsize \small { \sl 
$^{b}$ CP$^\mathbf 3$-Origins and DIAS, Southern Denmark University, \\ \vspace{-.2cm}Campusvej 55, DK-5230 Odense M, Denmark}}\\

\end{center}
\vspace{.8cm}

\setcounter{page}{1}

\begin{abstract}\baselineskip=15pt
 The measurements of the Higgs boson and top quark masses can be used to extrapolate the
 Standard Model Higgs potential at energies up to the Planck scale. Adopting a NNLO renormalization procedure, we:
i) find that electroweak vacuum stability is at present allowed, discuss the associated theoretical and experimental errors 
and the prospects for its future tests;
ii) determine the boundary conditions allowing for the existence of a shallow false minimum slightly below the Planck scale,
which is a stable configuration that might have been relevant for primordial inflation;
iii) derive a conservative upper bound on type I seesaw right-handed neutrino masses, following from the 
requirement of electroweak vacuum stability.
\end{abstract}

\maketitle 

%\end{titlepage}

\baselineskip=16pt

\vspace{.8cm}

%%%%%%%%%%%%%%%%%%%%%%%%%%%%%%%%%
\section{Introduction}

The recent discovery of a particle consistent with the Standard Model (SM) Higgs boson, announced by the ATLAS \cite{:2012gk} 
and CMS \cite{:2012gu} collaborations at CERN, is a milestone in particle physics; 
adding in quadrature statistical and systematic errors,
the mass of the particle turns out to be in the range $124.8 - 126.5$  GeV at $2\sigma$.

Here we assume that the new particle is actually the SM Higgs boson and study the implications that its mass value, 
together with other relevant parameters such as the top quark mass and the strong gauge coupling, has on the behavior of the Higgs potential at very 
high energy scales and, in particular, for the sake of electroweak vacuum stability. 

The project of extrapolating the Higgs potential up to the Planck scale is a long standing one \cite{1group,2group,3group},
and was revamped in the fall of 2011\,\cite{Holthausen:2011aa,Masina:2011aa,EliasMiro:2011aa,Xing:2011aa} after the first 
LHC hints of a Higgs boson were reported\,\cite{HCP11}. Recently, the tools for a Next-to-Next-to-Leading Order (NNLO) renormalization procedure 
were derived \cite{Mihaila:2012fm,Chetyrkin:2012rz,Bezrukov:2012sa,Degrassi:2012ry}.
So, there are now all the ingredients necessary to carry out this long standing project.  
Clearly, the extrapolation is based on the assumption that
there is a desert up to the Planck scale or, better, that possible new physics do not significantly affect the running of the Higgs quartic
coupling, which dominates the Higgs potential at high energy.

It is interesting that the recently discovered experimental Higgs mass range, combined with the experimental top mass
range, indicates a particularly intriguing high energy behavior of the Higgs potential, close to the transition between electroweak vacuum 
stability and metastability. 
This is due to the fact that,  for these Higgs and top mass values, the Higgs quartic coupling
can be very small or even negative.
Since the dependence on the top mass is strong and quite subtle, it is not surprising that different groups slightly disagree in the interpretation
of the results, some of them favoring \cite{Bezrukov:2012sa} and some others disfavoring \cite{Degrassi:2012ry} electroweak vacuum stability.

Traditionally the top pole mass was used in the analysis; however it has been pointed out \cite{Alekhin:2012py} that the top pole mass 
value used in previous analyses and taken to be the one measured at the Tevatron, $m_t^{exp} = 173.2 \pm 0.9$ GeV \cite{Lancaster:2011wr}, 
is not unambiguously derived and that a more careful derivation 
should be based instead on the running top mass in the $\overline{\rm MS}$ scheme, $\overline{m_t}(m_t)=163.3\pm2.7$ GeV. 
As was shown in \cite{Alekhin:2012py}, the top pole mass range consistently derived from the running one, $m_t= 173.3 \pm 2.8$ GeV, 
is plagued by a larger error than the Tevatron  measurement considered in \cite{Degrassi:2012ry}, 
rescuing  electroweak vacuum stability. 
 
In our analysis we keep as a free parameter the running top mass, rather than the pole one. In this way
we completely avoid the theoretical uncertainties associated with the top Yukawa matching procedure. As we are going to
discuss, the theoretical error associated with the Higgs quartic coupling matching \cite{Bezrukov:2012sa,Degrassi:2012ry}
turns out to be smaller than the one induced by the experimental uncertainty in the strong gauge coupling, $\alpha_3(m_Z)$.
Given the above mentioned range for the running top mass \cite{Alekhin:2012py}, 
we find that electroweak stability is allowed in the whole Higgs mass range \cite{:2012gk,:2012gu}. 
Stability  could soon be excluded if values of the running top mass $\overline{m_t}(m_t)<163$ GeV 
are excluded by the LHC.
Otherwise, testing electroweak vacuum stability would become very challenging, 
since this would require precision measurements of the Higgs and top masses, and also of $\alpha_3(m_Z)$.

A stable Higgs potential configuration which deserves particular interest is a shallow false minimum close to the Planck scale,
which could have been relevant for primordial inflation\,\cite{Masina:2011aa,Masina:2011un,Masina:2012yd}.
We show that such a configuration is realized only if the Higgs quartic coupling and its derivative satisfy very 
specific boundary conditions, possibly having a deep origin in quantum gravity.

As is well know, new physics in addition to the SM is required to explain the neutrino masses and mixings, as well as dark matter.
The mechanism responsible for the neutrino masses could affect the Higgs quartic coupling; 
as an example, we consider the impact that the inclusion of neutrino masses via a type I seesaw has on electroweak stability,
discussing in some detail the shallow false minimum configuration.

The paper is organized as follows. In sec.\,\ref{sec-1} we discuss the input parameters
and the NNLO renormalization procedure used to extrapolate the Higgs potential up to the Planck scale.
An analysis of electroweak vacuum stability and the associated constraints on the top and Higgs masses, 
with a detailed discussion of the theoretical errors and the prospects for the future, are presented in sec.\,\ref{sec-2}. 
In sec.\,\ref{sec-3} we investigate the 
boundary conditions leading to the particularly interesting configuration of a shallow false minimum below the Planck scale. 
Sec.\,\ref{sec-4} is devoted to the upper bound on the seesaw right-handed
neutrino masses following from the requirement of electroweak vacuum stability.
Conclusions are drawn in sec.\,\ref{sec-concl}. Appendix\,\ref{app-a} contains the relevant formulas for the NNLO running procedure
in the SM and, and those to incorporate the type I seesaw mechanism are found in appendix\,\ref{app-b}.

%%%%%%%%%%%%%%%

\section{Input parameters and renormalization at NNLO }
\label{sec-1}

The normalization of the Higgs quartic coupling $\lambda$ is chosen in this paper so that the potential for the physical 
Higgs $\phi_H$ contained in the Higgs doublet $H=(0,(\phi_H+v)/\sqrt{2})$ is given, at tree level, by
\begin{equation}
V(\phi_H)=\frac{\lambda}{6} \left( |H|^2 - \frac{v^2}{2} \right)^2 \approx  \frac{\lambda}{24}  \phi_H^4  \,\,\,\, ,
\label{eq-V}
\end{equation}
where $v=1/(\sqrt2 G_\mu)^{1/2}=246.221{\rm ~GeV} $ and $G_\mu=1.1663787(6)\times 10^{-5}/$GeV$^{2}$ 
is the Fermi constant from muon decay  ~\cite{PDG}. 
The approximation in eq.\,(\ref{eq-V}) holds when considering large field values. 
According to our normalization, the physical Higgs mass satisfies the tree level relation $m_H^2= \lambda v^2 /3$.
In addition, the mass of the fermion $f$ reads, at tree level,  $ m_f =   h_f  v/\sqrt{2}$,
where $h_f$ denotes the associated Yukawa coupling.

In order to extrapolate the behavior of the Higgs potential at very high energies, we adopt the $\overline{\rm MS}$ scheme 
and consider the Renormalization Group (RG) evolution for the relevant couplings which, in addition 
to the Higgs quartic coupling $\lambda$, are the gauge $g$, $g'$, $g_3$, and the top Yukawa $h_t$ couplings. 
We work at NNLO, namely 3-loops for the $\beta$-functions and 2-loops for the matching conditions at some suitable scale.

It is customary to introduce the dimensionless parameter $t=\log \mu/m_Z$, where $\mu$ stands for the renormalization scale 
and $m_Z$ is the $Z$ boson mass.
The RG equations for the relevant couplings are then given by
\begin{eqnarray}
\nonumber \frac{d}{d t} \lambda(t)&=&\kappa \beta_\lambda^{(1)}+\kappa^2  \beta_\lambda^{(2)} +\kappa^3  \beta_\lambda^{(3)} \,,  \\  
\nonumber
\frac{d}{d t}h_t(t)&=&\kappa\beta_{h_t}^{(1)}+ \kappa^2  \beta_{h_t}^{(2)}+ \kappa^3 \beta_{h_t}^{(3)}\,,\\  
\nonumber
\frac{d}{dt} g(t)&=&\kappa\beta_g^{(1)}+\kappa^2   \beta_{g}^{(2)} +\kappa^3   \beta_{g}^{(3)}\,,\\  
\nonumber
\frac{d}{dt} g'(t)&=&\kappa\beta_{g'}^{(1)}+ \kappa^2  \beta_{g'}^{(2)}+ \kappa^3  \beta_{g'}^{(3)}\,,\\  
\frac{d}{dt} g_3(t)&=&\kappa\beta_{g_3}^{(1)}+ \kappa^2  \beta_{g_3}^{(2)}+ \kappa^3 \beta_{g_3}^{(3)} \,,\, 
%+ \kappa^4 \beta_{g_3}^{(4)}\,, 
\label{eq-RGE}
\end{eqnarray}
where $\kappa = 1/(16 \pi^2)$ and the apex on the $\beta$-functions represents the loop order.
The 1-loop and 2-loop expressions for the $\beta$-functions can be found 
{\it e.g.} in ref.\,\cite{FJSE} (see also \cite{GWP,altri, LV,vRVL,LX,Czakon:2004bu}).
Recently, the complete 3-loop $\beta$-functions for all the SM gauge couplings have been presented by Mihaila, Salomon and Steinhauser 
in ref.\,\cite{Mihaila:2012fm}, while the leading 3-loop terms in the RG evolution of $\lambda$, $h_t$
and the Higgs anomalous dimension have been computed by Chetyrkin and Zoller in ref.\,\cite{Chetyrkin:2012rz}.
For the sake of completeness, the expressions for the $\beta$-functions up to 3-loops are collected in appendix A.

The matching of the running gauge couplings
is done at the $Z$ boson pole mass\footnote{We use the value of the strong coupling at $m_Z$ and apply immediately the six flavor running.
The correction that would result by running with five flavors up to the top mass is very small and can be neglected, as discussed 
in \cite{Bezrukov:2012sa}.} ,
$m_Z$. 
The numerical values used for the related $\overline{\rm MS}$ observables are taken from the latest Particle Data Group SM fit results~\cite{PDG}:
\begin{eqnarray}
\alpha_{em}^{-1}(m_Z)=127.944 \pm 0.014 \,\,, \,\, \alpha_3(m_Z) = 0.1196 \pm 0.0017 \,\, , \\
\,\,\sin^2\theta_W(m_Z)  = 0.23116 \pm 0.00012 \,\,,  \,\,\,m_Z=91.1874 \pm 0.0021~\textrm{GeV} \,\,. \nonumber \label{eq-3}
 %\,\, m_W=80.381 \pm 0.014~\textrm{GeV}  \,\,.\,\,\,\,\,\,\,\,\,\,\, \,\,\,\,\,\,\,\,\,\,\,\,\,\,\,\, \,\,\,\,\,\,\,\,\,
\end{eqnarray}

To match the $\overline{\rm {MS}}$ running quartic coupling $\lambda(\mu)$ %as a function of the renormalization scale  $\mu$
with the Higgs pole mass $m_H$ is more complicated and requires one to exploit an expansion, 
\begin{equation}
\lambda(\mu)= \sum_{n=1,2,3,...} \lambda^{(n)}(\mu) \, =\, 3 %\sqrt{2} G_\mu 
\frac{m_H^2}{v^2} \Bigl(1+ \delta^{(1)}_H(\mu)   + \delta^{(2)}_H(\mu) +\dots \Bigr)\ ,
\label{match-la}
\end{equation}
which is known at present at NLO: $\delta^{(1)}_H(\mu)$ is the 1-loop  ${\cal O}(\alpha)$ result of Sirlin and Zucchini~\cite{SZ} 
while $\delta^{(2)}_H(\mu)$ is the recently calculated  2-loop result, composed of a QCD contribution of ${\cal O}(\alpha \alpha_3)$ 
\cite{Bezrukov:2012sa,Degrassi:2012ry} and a Yukawa contribution  \cite{Degrassi:2012ry}. More details can be found in appendix A.
As is well known, there is some arbitrariness in the choice of the matching scale $\mu$ in eq.\,(\ref{match-la}), which introduces a "theoretical" error
in the RG procedure. 
In this work, we choose to perform the matching of the Higgs quartic coupling $\lambda$ at the scale $\mu=m_H$.
The theoretical uncertainty is estimated by performing the matching also at different scales
and by evolving $\lambda$ via RG running until $\mu=m_H$. The spread in the numerical values obtained for $\lambda(m_H)$
can then be used to infer the magnitude of the theoretical error. 

\begin{figure}[h!]
\vskip .5cm 
 \begin{center}
\includegraphics[width=10cm]{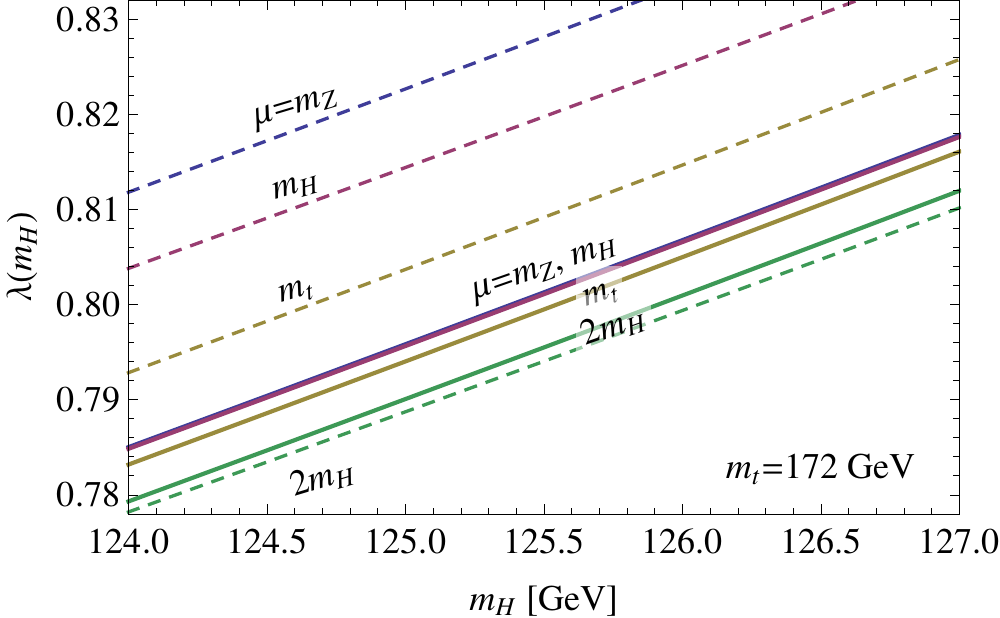} 
 \end{center}
\caption{\baselineskip=15 pt
Value of $\lambda(m_H)$ obtained by performing the matching at different scales $\mu$, indicated by the labels, as a function of $m_H$. 
The solid (dashed) lines are obtained by including corrections up to 2-loop (1-loop).  We fixed $m_t=172$ GeV 
(for different values see eq.(\ref{eq-lmh})). 
}
\label{fig-match-la}
\vskip .1 cm
\end{figure}

This is illustrated in fig.\,\ref{fig-match-la}, assuming for definiteness a top pole mass $m_t=172$\,GeV.
The dashed and solid curves show the value of $\lambda(m_H)$ obtained by including the corrections up to 1-loop and 2-loop respectively,
for various choices of the matching scale: from top to bottom $\mu=m_Z$, $m_H$, $m_t$, $2 m_H$.
One can see that, working at the 1-loop, the theoretical uncertainty is about $5 \%$.
The inclusion of the 2-loop corrections given in ref.\,\cite{Degrassi:2012ry} reduces the theoretical 
uncertainty down to about $0.7 \%$.
Notice also that the preferred region shrinks  to small $\lambda$ values and that $\mu=m_Z$ and $\mu=m_H$ 
nearly overlap. 
More generally, one can use the following expression for the 2-loop result,
\begin{equation}
\lambda(m_H)=0.8065 + 0.0109\, (m_H [{\rm GeV}]- 126) + 0.0015 \, (m_t[{\rm GeV}] - 172) \, _{-  0.0060}^{ + 0.0002} \,\, ,
\label{eq-lmh}
\end{equation}
where the mean value refers to $\mu=m_H$.
The reference values of $m_H $ and $m_t$ used 
in eq.(\ref{eq-lmh}) are not the central values that will be used in the following analysis; they are just "round numbers" 
allowing for an easy inspection of the variation of $\lambda(m_H)$ as a function of $m_H$ and $m_t$. 

Notice that it is not possible to compare directly eq.(\ref{eq-lmh}) with eq.(63) of 
ref.\cite{Degrassi:2012ry}, where $\lambda(m_t)$ is rather displayed (adopting a normalization
differing from ours by a factor of $6$ and choosing as reference values for  $m_H$ and $m_t$ their central ones):
$\lambda(m_t)= 0.12577 +0.00205\,  (m_H [{\rm GeV}] -125) -0.0004 \, (m_t [{\rm GeV}]-173.15) \pm 0.00140$,
where the error is obtained by varying the matching scale between $m_Z$, $m_t$,  $2 m_t$ \cite{Degrassi:2012ry}.
We checked that our numerical code gives a result for $\lambda(m_t)$ consistent with the one of 
eq.(63) of ref.\cite{Degrassi:2012ry}. Indeed, choosing $m_H=125$ GeV and $m_t= 173.15$ GeV, our code gives 
$\lambda(m_t) /6=0.12605, 0.12575, 0.12412=0.12575^{+.0003}_{-.0016}$, when the matching scale 
$\mu=m_Z, m_t, 2 m_t$ respectively. This shows that the two results perfectly agree  for $\mu=m_t$, that
the lower errors  (associated to the difference between $\mu=m_t$ and $\mu=2 m_t$) are in substantial agreement,
while the upper errors  (associated to the difference between $\mu=m_Z$ and $\mu= m_t$) are slightly different, 
ours being smaller.

\begin{figure}[t!]
 \begin{center}
 \vskip .5 cm
 \includegraphics[width=10cm]{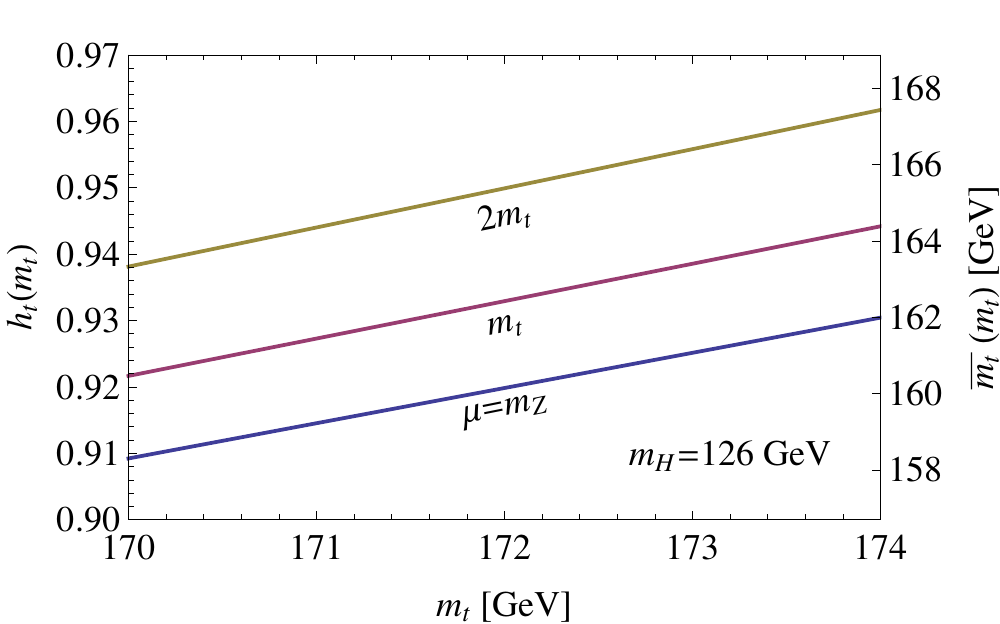} 
 \end{center}
\caption{\baselineskip=15 pt
Values of $h_t(m_t)$ and $\overline{m_t}(m_t)$ as a function of $m_t$. The curves are obtained by matching at different scales, which are indicated by the labels. 
We fixed $m_H=126$ GeV for definiteness but the results do not significantly dependent on $m_H$, provided it is chosen 
in its experimental range.}
\label{fig-match-h}
\vskip .1 cm
\end{figure}

It is common to extrapolate the $\overline{\textrm{MS}}$  top Yukawa coupling $h_t(\mu)$ from the matching condition between the running 
top mass $\overline{m_t}(\mu) $ and the top pole mass $m_t$:
\begin{equation}
 h_t(\mu) \frac{v}{\sqrt{2}}=\overline{m_t} (\mu)= m_t  \Bigl(1+  \delta_t(\mu)\Bigr)\,\, , 
\,\, \, \delta_t(\mu)=\delta_t^{W}(\mu)+\delta_t^{QED}(\mu)+\delta_t^{QCD}(\mu)\, ,  \label{eq-match-h}
\end{equation}
where $\delta_t^W+ \delta_t^{QED}$ represent the electroweak contribution, which is known at 1-loop \cite{HK},
while $\delta_t^{QCD}$ is the QCD one. The QCD 1-loop result is known since many years \cite{HK}; the QCD 2-loop and 3-loop results
as a function of the matching scale $\mu$ are given in \cite{Chetyrkin:1999qi} (see also \cite{GBGS,BGS,FJTV,Melnikov:2000qh,JK}).
%{\color{red} Ci sarebbe anche il contributo misto che non ho messo perche' troppo piccolo e noto (credo) solo a mt: 
%mixed two-loop strong/weak~\cite%{JK} contributions. }
The matching is usually done at the top pole mass scale, and the theoretical error associated to the arbitrariness of the matching scale 
can be estimated as before, namely by comparing the values of $h_t(m_t)$ obtained with different matching scales.  
This is represented in fig. \ref{fig-match-h}, where the curves are obtained by working at 2-loop and using, from bottom to top,  
$\mu=m_Z,m_t,2 m_t$.
The plot shows that the associated  theoretical uncertainty is about $2 \%$. The analytical expression for $h_t(m_t)$ is: 
\begin{equation}
h_t(m_t)=0.933  + 0.006\, (m_t [{\rm GeV}]- 172)\, ^{+0.017}_{- 0.013} \label{eq-fit-h}  \,\, .
\end{equation}
%which is also equivalent to $\overline{m_t}(m_t)[{\rm GeV}]=m_t[{\rm GeV}]-9.6 \pm 2.5 $.
The variations of $h(m_t)$ due to the experimental range of $\alpha_s$ and $m_H$ have not been explicitly 
written in eq.(\ref{eq-fit-h}) because they are negligible (respectively of order $10 \%$ and $1\%$) with respect to 
the variation of $h(m_t)$ due to the experimental range of $m_t$. The error quoted in eq.(\ref{eq-fit-h}) then refers 
only to the theoretical error coming from varying the matching scale $\mu$ from $m_Z$, $m_t$ (mean value), $2 m_t$. 
Notice that our result perfectly agrees with the analogous expression derived in ref. \cite{Degrassi:2012ry}, where
however the error due to the variation of the matching was not estimated.

The procedure adopted in previous analyses of the stability of the electroweak vacuum, including the latest
ones \cite{Bezrukov:2012sa,Degrassi:2012ry}, 
was to use the experimental value of $m_t$, identified with the one measured at the Tevatron 
by the CDF and D0 collaborations, $m_t^{exp}=173.2 \pm 0.9$\,GeV\,\cite{Lancaster:2011wr}, 
%(and to be measured at the LHC),
to extrapolate the running Yukawa $h_t(m_t)$ via eq.\,(\ref{eq-fit-h}).
However,  as discussed in ref.\,\cite{Alekhin:2012py}, it is not meaningful to use the mass parameter 
provided by the Tevatron as the pole top mass to be inserted in eq.\,(\ref{eq-fit-h}):
the running top mass in the $\overline{\rm MS}$ scheme
is instead a well defined parameter that can be directly extracted at NNLO from Tevatron measurements 
of the inclusive top pair production cross-section,
giving  $\overline{m_t} (m_t)=163.3 \pm 2.7$ GeV\,\cite{Alekhin:2012py}. 
So, it is conceptually more robust and practically more convenient to extract the top Yukawa coupling directly from  $\overline{m_t} (m_t)$,
%simply by means of eq.\,(\ref{eq-match-h}), 
as will be done in the following\footnote{At difference, ref.\,\cite{Alekhin:2012py} proceeds in
a more complicated way: the value of $\overline{m_t} (m_t)$ is translated into a value of $m_t$, to be inserted in the 
%analog of our eq.(\ref{eq-fit-h}) 
expression of the lower bound on $m_H$ ensuring electroweak vacuum stability as derived in ref.\,\cite{Degrassi:2012ry}.}.
Our results will thus be presented as a function of $\overline{m_t} (m_t)$. 

Notice that, according to eq.\,(\ref{eq-fit-h}), the value of the top pole mass can be easily recovered via the relation
$m_t=\overline{m_t}(m_t)+9.6 \,^{+2.9}_{-2.3} $ GeV, which however is plagued by a large uncertainty. 
In ref.\,\cite{Alekhin:2012py}  it was found that, by doing a scheme transformation to NNLO accuracy from the running to the pole top mass, 
the  range $\overline{m_t} (m_t)=163.3 \pm 2.7$ GeV is equivalent to $m_t= 173.3 \pm 2.8$ GeV. 
Hence, while displaying our results as a function of $\overline{m_t} (m_t)$ as already stated, motivated by the results of 
ref.\,\cite{Alekhin:2012py}, in some plots (as the one in fig.\,\ref{fig-line}) we will link the value of the top pole mass to the running mass 
via the simple relation $m_t=\overline{m_t}(m_t)+10 $ GeV.

Before presenting the results of our analysis in the following sections, we recall that, in order to carefully study the shape of the Higgs 
potential at high energy, one should consider the renormalization improved effective potential.
This can be done by introducing an effective coupling, $\lambda_{eff}(\mu)=\lambda(\mu) +\Delta \lambda(\mu)$, so that
\begin{equation}
V_{eff}(\phi_H)= \frac{\lambda_{eff}(\mu)}{24} \phi_H^4 \,\,\,.
\end{equation} 
The expression for $\Delta \lambda(\mu)$ is known up to 2-loop \cite{FJSE,2group} (and given, for instance, in\,\cite{Degrassi:2012ry}).
Since the scalar contribution is not well defined when $\lambda$ is negative (a logarithm of a negative quantity appears), 
in the following we consider the renormalization improved potential at the tree level, and identify $\mu$ with $\phi_H$. 
It is well known that this simplification has a negligible impact in the determination of the vacuum stability bound 
(for a detailed discussion see {\it e.g.} ref.\,\cite{Bezrukov:2012sa}) to be discussed in the next section.

%We will also take $m_b=4.2~\textrm{GeV}$.  ... discutere meglio questo....

%%%%%%%%%%%%%%%%%%%%%%%%%%%%%%%%%%%%%%%%%%%%%%%%%%%%%

\section{Electroweak Vacuum Stability}
\label{sec-2}

The experimental region of the values of the Higgs and top masses is very intriguing from the theoretical point of view,
since the Higgs quartic coupling could be rather small, vanish or even turn negative at a scale slightly smaller than the Planck scale. 
Accordingly, the behavior of the Higgs potential at high energy changes drastically:
if $\lambda(\mu)$ is always positive, the electroweak vacuum is a global minimum, possibly accompanied by another local minimum 
just below the Planck scale, which could have played a role in primordial inflation\,\cite{Masina:2011un,Masina:2011aa,Masina:2012yd}; 
if $\lambda(\mu)$ turns negative below $M_{\rm Pl}$, the electroweak vacuum correspondingly becomes metastable \cite{2group,3group}.

\begin{figure}[b!]
 \begin{center}
 \vskip .8 cm
\includegraphics[width=7.1cm]{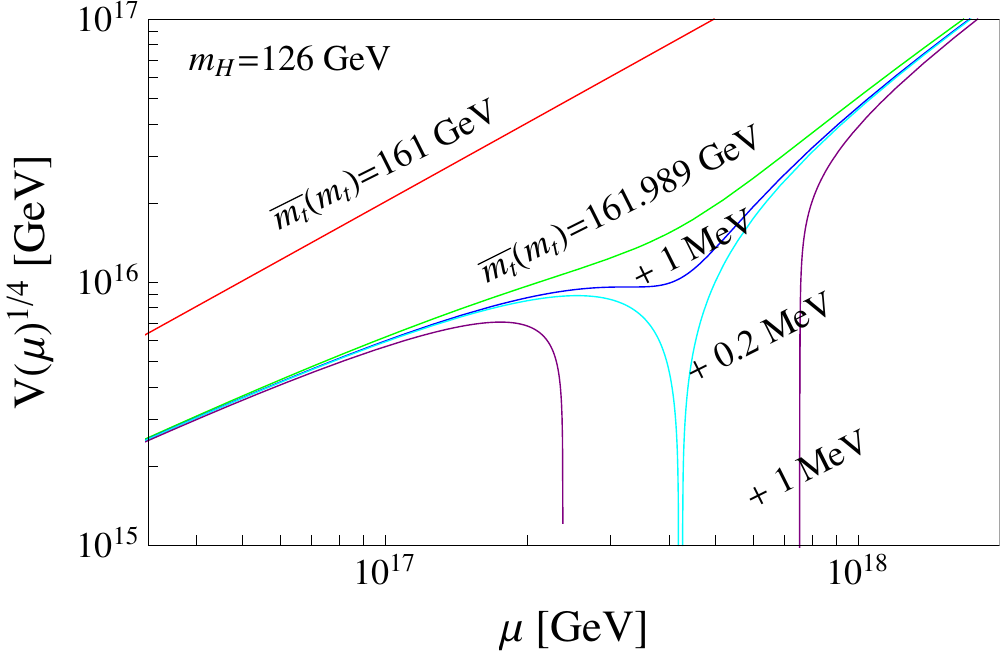}\,\,\,\,\,\,\,  \, \includegraphics[width=7.5cm]{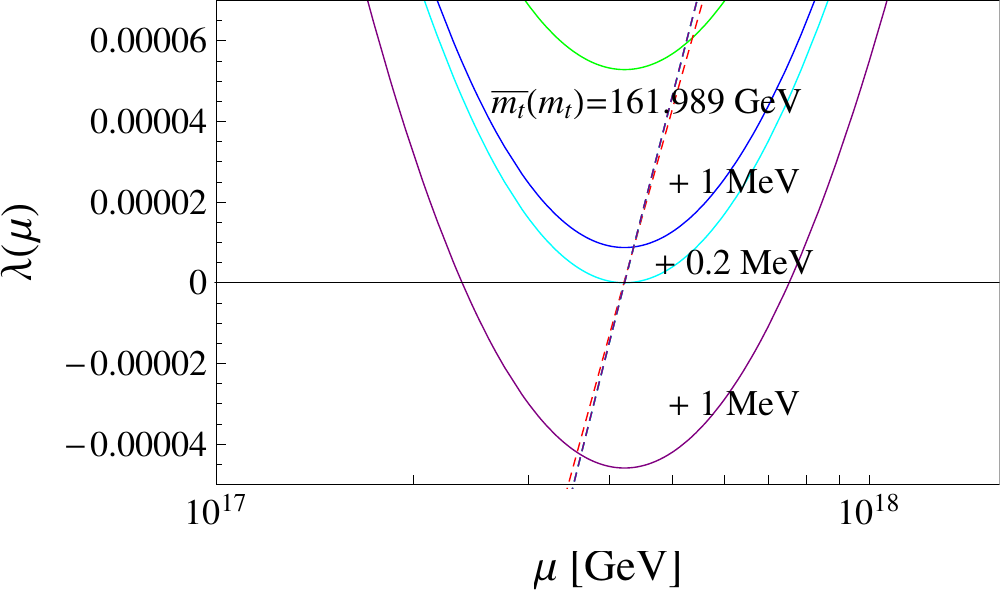}
\end{center}
\caption{\baselineskip=15 pt
The SM Higgs potential (left) and the quartic Higgs coupling (right) as functions of the renormalization scale $\mu$, for $m_H=126$ GeV 
and  different values of $\overline{m_t}(m_t)$, increasing from top to bottom by the amount indicated by the labels.  
The dashed curve in the right plot shows the associated value of $\beta_\lambda(\mu)$.
The other input parameters are fixed at the central values discussed in the previous section. }
\label{fig-shapes}
\vskip .1 cm
\end{figure}

These drastically different possibilities for the behavior of the renormalization improved Higgs potential at high energy
are illustrated in the left plot  fig.\,\ref{fig-shapes}, where $m_H=126$\,GeV and some specific values for $\overline{m_t}(m_t)$ have been selected,
increasing from top to bottom. The right plot shows the associated values of $\lambda(\mu)$. 
Let start considering the value $\overline{m_t}(m_t)=161.989$ GeV. Increasing the latter  by just $1$ MeV, the potential 
develops an inflection point; notice that the associated $\lambda(\mu)$ becomes as small as ${\cal O}(10^{-5})$.
%this happens for $\lambda$ has a minimum 
%at a scale $\mu_\beta$ and if $\lambda(\mu_\beta) = 8.7 \times 10^{-6}$. 
%Such a special value could correspond to boundary conditions dictated by an eventual merging with quantum gravity  \cite{Holthausen:2011aa}
%(more on this later). 
Increasing again $\overline{m_t}(m_t)$ by about $200$ keV, the minimum of $\lambda(\mu)$ is equal to zero: 
a second vacuum degenerate 
with the electroweak one is obtained. 
Further increasing $\overline{m_t}(m_t)$ makes  $\lambda(\mu)$ turn negative:  the electroweak vacuum becomes  metastable. 

The dashed curve in the right plot in fig.\,\ref{fig-shapes} shows the evolution of $\beta_\lambda(\mu)=d \lambda(\mu)/dt$ for the same
parameter values; there is only a single dashed curve because $\beta_\lambda(\mu)$ mildly depends on $\overline{m_t}(m_t)$
if the latter is in the range $161-163$ GeV. 
Let call $\mu_\beta$ the renormalization scale such that $\beta_\lambda(\mu_\beta)=0$. 
Clearly,  only in the case of two degenerate vacua the conditions   $\beta_\lambda(\mu_\beta)=0$ and $\lambda(\mu_\beta)=0$ are 
simultaneously met. For a shallow false minimum we instead have $\beta_\lambda(\mu_\beta)=0$ and $\lambda(\mu_\beta)={\cal O}(10^{-5})$,
as already mentioned. 

\begin{figure}[h!]
 \begin{center}\vskip 0.5cm
 \includegraphics[width=9.5cm]{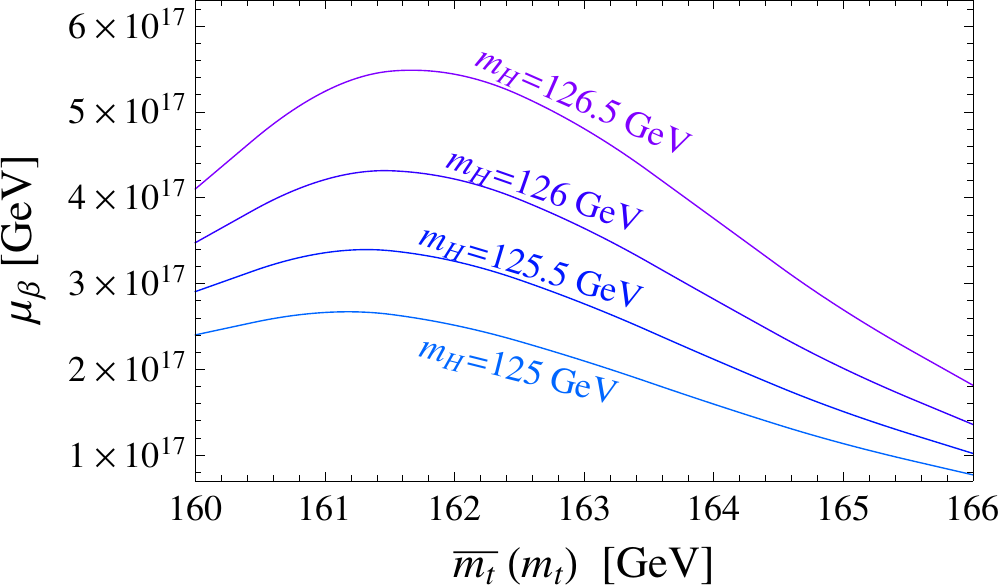}
\end{center}
\caption{\baselineskip=15 pt
The scale $\mu_\beta$ as a function of  $\overline{m_t}(m_t)$ and for different values of $m_H$,
as indicated by the labels.}
\label{fig-mubmt}
\vskip .5 cm
\end{figure}

In fig. \ref{fig-mubmt} we show how $\mu_\beta$ depends on $\overline{m_t}(m_t)$, 
for various values of $m_H$. It is interesting that $\mu_\beta$ is maximized and nearly constant
for the values of $\overline{m_t}(m_t)$ for which $\lambda(\mu)$ is very small.

%The value of  $\mu_\beta$ increases with $m_H$, as shown in fig. \ref{fig-mu}.

\begin{figure}[h!]
 \begin{center}
\includegraphics[width=11.5cm]{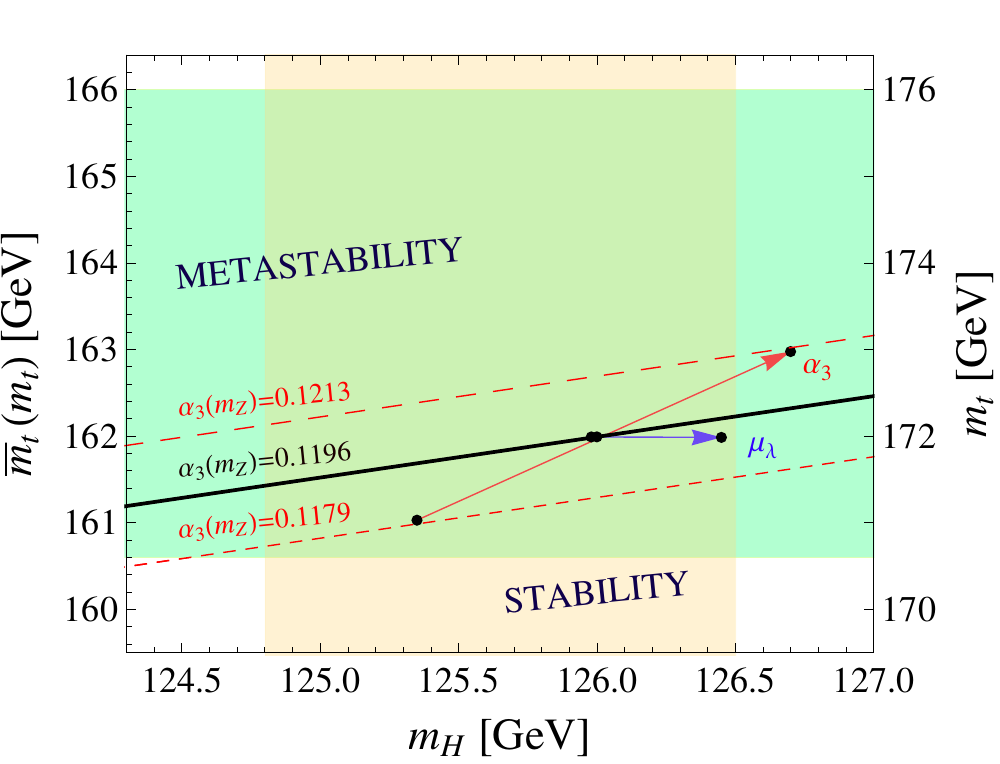} 
 \end{center}
\caption{\baselineskip=15 pt
The solid (black) line marks the points in the plane $[m_H,\overline{m_t}(m_t)]$ where a second vacuum, 
degenerate with the electroweak one, is obtained just below
the Planck scale. The (red) diagonal arrow shows the effect of varying $\alpha_3(m_Z)=0.1196\pm 0.0017$ \cite{PDG}; 
the (blue) horizontal one shows the effect of varying $\mu_\lambda$ (the matching scale of $\lambda$) from $m_Z$ up to $2 m_H$.  
The shaded (yellow) vertical region is the $2\sigma$ ATLAS\,\cite{:2012gk} and CMS\,\cite{:2012gu}
combined range,  $m_H=125.65 \pm 0.85$ GeV;
the shaded (green) horizontal region is the range $\overline{m_t} (m_t)=163.3 \pm 2.7$\,GeV, equivalent to 
$m_t=173.3\pm2.8$ GeV\,\cite{Alekhin:2012py}. }
\label{fig-line}
\vskip .5 cm
\end{figure}

We now turn to the determination of the points in the  plane $[m_H,\overline{m_t}(m_t)]$ allowing for the existence of a second minimum 
degenerate with the electroweak one.
These points belong to a line separating the stability from the metastability region, see fig.\,\ref{fig-line}:
in the lower part of the plot $\lambda(\mu)$ is always positive, while in the upper part it becomes negative before reaching the Planck scale. 
The configuration of a shallow false minimum belongs to the stability region, but the associated points are so close to the transition line 
that they could not be distinguished visually.

The transition line of fig.\,\ref{fig-line} was obtained with the input parameter values discussed in the previous section
and by matching the running Higgs quartic coupling at $m_H$.
Clearly, it is also important to estimate the theoretical error associated to experimental ranges of the input parameters 
and the one associated to the matching procedure.
To illustrate this, we consider in particular the point on the transition line associated to the value $m_H=126$ GeV;
for such point, $\lambda$ and $\beta_\lambda$ both vanish at a certain scale $\mu_\beta$ (see fig.\,\ref{fig-mubmt}).
The arrows show how, if some inputs or the matching scale are changed, the position of this point 
have to change in order to keep having, at the same scale $\mu_\beta$, a vacuum degenerate with the electroweak one. 
%As for the input parameters, the error is largely dominated by $\alpha_3(m_Z)$.
The diagonal arrow is obtained by varying the strong coupling in its allowed range, $\alpha_3(m_Z)=0.1196 \pm 0.0017$ \cite{PDG}; 
the short (long) dashed line shows how the solid line would move if $\alpha_3(m_Z)$ were equal to its minimum (maximum) 
presently allowed value. 
Notice that the error on $\alpha_3(m_Z)$ induces an uncertainty in both the Higgs and top masses of about $\pm 0.7$ GeV.
In ref.\cite{Degrassi:2012ry} the impact of the variation of  $\alpha_3(m_Z)$ on $m_H$ was estimated to be $\pm 0.5$ GeV 
(see their table 1).
The two results are in substantial agreement, considering that in our analysis $\alpha_3(m_Z)=0.1196 \pm 0.0017$ at $1\sigma$
\cite{PDG},
while ref.\cite{Degrassi:2012ry} considers a smaller error,  $\alpha_3(m_Z)=0.1184 \pm 0.0007$ at $1\sigma$.
Since the variation of the other input parameters in eq.(\ref{eq-3}) induces much smaller effects then the one due to $\alpha_3(m_Z)$, 
they have not been reported in the fig.\ref{fig-line}.
The horizontal arrow represents instead the theoretical error obtained by varying $\mu_\lambda$, the matching scale of the Higgs 
quartic coupling,  
from $\mu=m_Z$ to  $\mu=2 m_H$; 
notice that the associated error is very asymmetric (see fig.\,\ref{fig-match-la}): essentially it can only 
enhance $m_H$, by at most $0.5$ GeV. Clearly, similar considerations apply to each point of the transition line. 
We note that in ref.\cite{Degrassi:2012ry} the impact of the variation of the matching scale of $\lambda$ on $m_H$ was estimated to be 
$\pm 0.7$ GeV (see their table 1), hence close to our estimate but with a symmetric error.

\begin{figure}[h!]
 \begin{center}
\includegraphics[width=11.5cm]{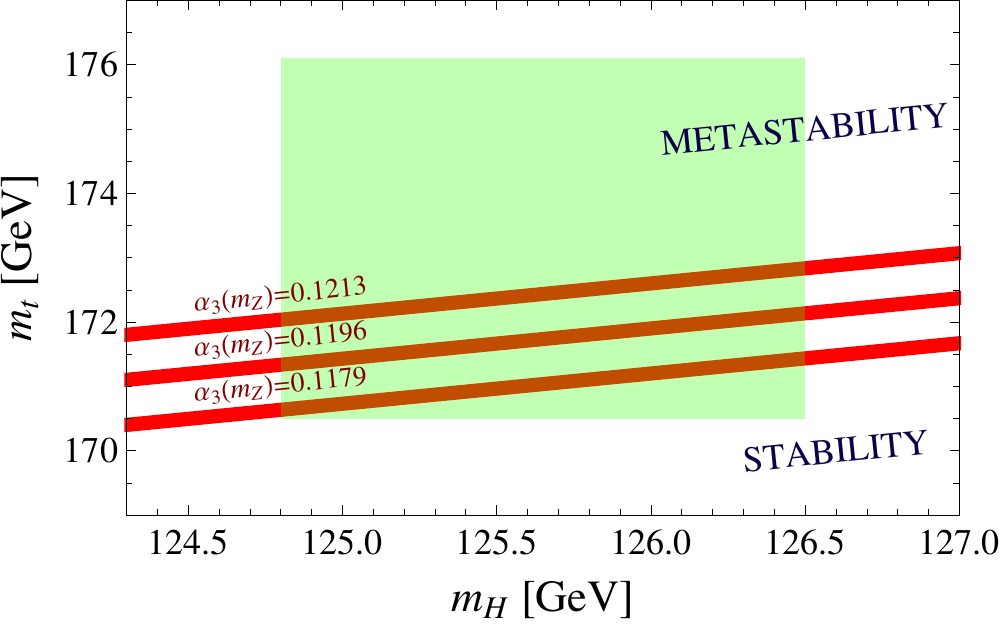} 
 \end{center}
\caption{\baselineskip=15 pt
The transition line between stability and metastability in the plane $[m_H,m_t]$ and
for fixed values of $\alpha_3(m_Z)=0.1196\pm 0.0017$ \cite{PDG}.  The thickness of the lines represents the
theoretical error due to the variation of $\mu_\lambda$ (the matching scale of $\lambda$) from $m_Z$ up to $2 m_H$.  
The shaded region is obtained by intersecting the $2\sigma$ ATLAS\,\cite{:2012gk} and CMS\,\cite{:2012gu}
combined range ($m_H=125.65 \pm 0.85$ GeV) with the running top mass range given by ref.\cite{Alekhin:2012py},
$\overline{m_t} (m_t)=163.3 \pm 2.7$\,GeV, equivalent to $m_t=173.3\pm2.8$ GeV\,. }
\label{fig-resI}
\vskip .5 cm
\end{figure}

Fig.\,\ref{fig-line} shows that stability can be achieved in the whole experimental range for $m_H$ (shaded vertical region), 
but this is not the case for $\overline{m_t}(m_t)$ (shaded horizontal region). 
So, it is convenient to write down the condition of 
electroweak vacuum stability under the form of an upper bound on the running top mass:
\begin{equation}
\overline{m_t}(m_t)[{\rm GeV}]  \le 162.0   + 0.47 \,(   m_H[{\rm GeV}] - 126   )  + 0.7 \, \left( \frac{\alpha_3(m_Z) - 0.1196}{0.0017} \right)
%\pm 0.7_{\alpha_3}
-0.2^{(\mu_\lambda)}_{th} \,\,\, ,
\label{eq-boundmt}
\end{equation}
where the last term accounts for the (very asymmetric) theoretical error induced by the matching of $\lambda$. The latter turns 
out to be smaller than the variation induced by varying $\alpha_3(m_Z)$ in its presently allowed experimental range.
We recall that the relation between the running and pole top mass is simply $m_t= \overline{m_t}(m_t)  +10$ GeV.
Fig.\,\ref{fig-resI} summarizes our results for the determination of the transition line between stability and metastability 
in the $[m_H,m_t]$ plane.
The three lines correspond to the central and $\pm 1\sigma$ values of $\alpha_3(m_Z)$ \cite{PDG} and their thickness 
represents the theoretical error due to the matching of $\lambda$. 
The shaded rectangle emphasizes the present allowed region  for $m_t$ \cite{Alekhin:2012py}
and $m_H$ \cite{:2012gk,:2012gu}.
According to our analysis it is not possible, given the present experimental situation, to understand whether we live in a stable or 
metastable vacuum configuration\footnote{Of course, assuming that the running of $\lambda$ happens as in the SM up to energies 
close to the Planck scale, without significant modifications.}.

In order to discriminate between the two possibilities, it would be crucial to better determine $\overline{m_t}(m_t)$. 
%, even more than $m_H$.
As discussed in \cite{Alekhin:2012py}, after LHC the Higgs mass will presumably be known with 
an accuracy of ${\cal O}(100)$ MeV \cite{futuremh}, 
but the precision on the top mass would improve only by a factor of two. 
For instance, if the whole range of $\overline{m_t}(m_t)<163$ GeV (or, equivalently $m_t<173$ GeV)
would be excluded, we would conclude that our vacuum is metastable; otherwise the investigations should continue.
 
A self-consistent and precise determination
of the top quark mass can best be performed at a high-energy electron-positron collider, with 
a planned accuracy of ${\cal O}(100)$ MeV. 
Moreover, at an electron-positron collider $\alpha_3(m_Z)$ could be determined with an accuracy close 
to or better than $\Delta\alpha_3(m_Z) = 0.0007$ (this precision is sometimes currently adopted\,\cite{Bezrukov:2012sa,Degrassi:2012ry}  
but cannot be considered to be conservative according to ref.\,\cite{Alekhin:2012py}).
At this stage, if the stability region will still have an overlap with the allowed ranges of the top and Higgs masses,  
we will be mostly limited by the theoretical uncertainty associated to $\mu_\lambda$. 
Notice also that it is not realistic to hope to distinguish the case of two degenerate minima with the one of a shallow false
minimum, since the difference in the top mass is just about $200$ keV (see fig. \ref{fig-shapes}).

\begin{figure}[t!]
 \begin{center}
\includegraphics[width=7.8cm]{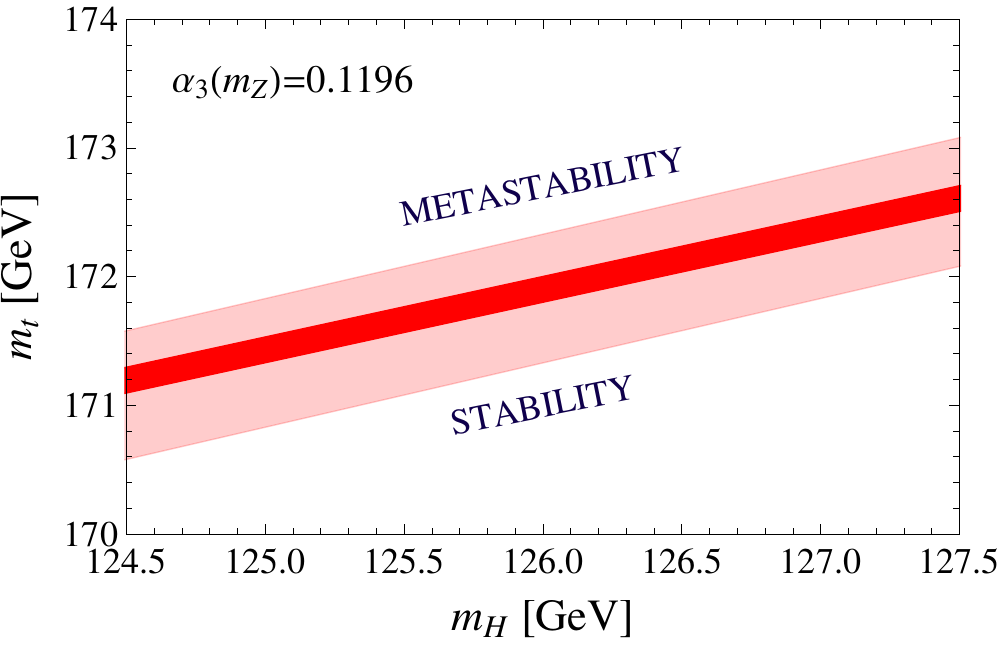}\,\,\,\, \includegraphics[width=7.8cm]{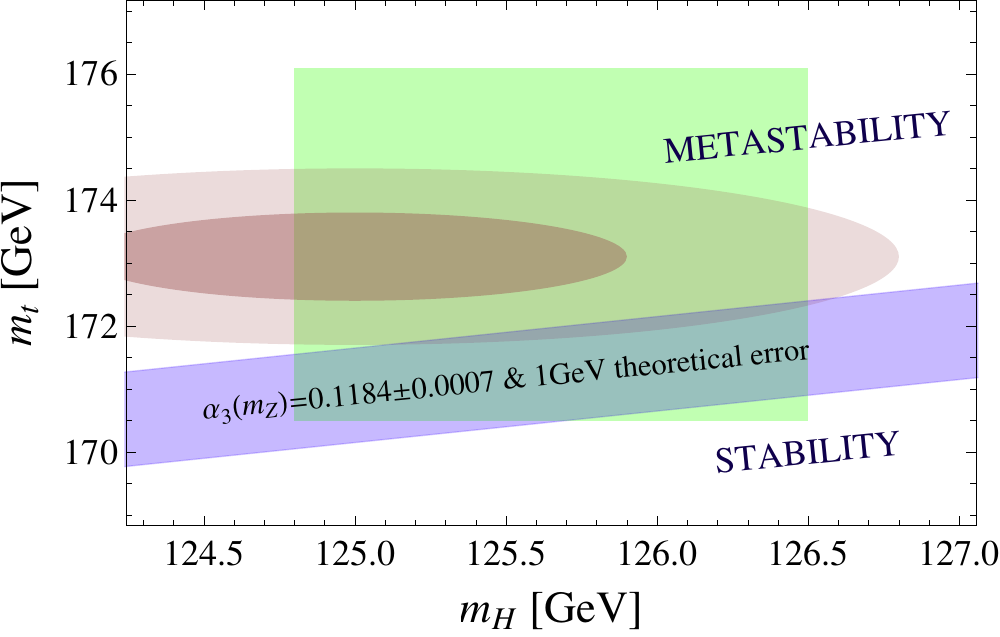} 
 \end{center}
\caption{\baselineskip=15 pt
Left: Theoretical uncertainty in the determination of the transition line between stability and metastability according to 
our eq.\,(\ref{eq-boundmt}) (thinner) and eq.\,(\ref{eq-14}) of ref.\,\cite{Degrassi:2012ry} (thicker). For definiteness 
we choose $\alpha_3(m_Z) = 0.1196$. Right: transition line between stability and metastability according to eq.\,(\ref{eq-14}) of 
ref.\,\cite{Degrassi:2012ry}; the thickness of the band accounts for both the $1$ GeV theoretical error and the experimental 
error due to the variation of $\alpha_3(m_Z)$ in the range $0.1184\pm0.0007$, as done in \cite{Degrassi:2012ry}. The (brown)
shaded disks represent the $1\sigma$ and $2\sigma$ combined ranges for $m_t$ and $m_H$ used in \cite{Degrassi:2012ry}
(see their fig.\,5). The (green) rectangle allows for the comparison with the ranges of $m_t$ and $m_H$ used here (see 
fig.\,\ref{fig-resI}).} 
\label{fig-compIS}
\vskip .5 cm
\end{figure}

We now discuss how to compare eq.\,(\ref{eq-boundmt}) and fig.\,\ref{fig-resI} with previous literature results, in particular those of 
ref.\,\cite{Degrassi:2012ry}, since the authors claimed that "absolute stability of the Higgs potential is excluded at $98\%$ C.L.
for $m_H<126$ GeV" (see also their fig.\,5).
In the latter work the stability condition is indeed expressed under the form of a lower bound on the Higgs mass: 
$m_H \ge 129.4 + 1.4\, (m_t-173.1)/0.7 -0.5\, (\alpha_3(m_Z)-0.1184)/0.0007 \pm 1_{th}$, where all masses are in GeV and the
last term represents the overall theoretical error.
Combining in quadrature their theoretical uncertainty and their experimental errors on $m_t$ and $\alpha_3(m_Z)$, the authors 
derive $m_H>129.4 \pm 1.8$ GeV, which motivates the quoted claim of ref.\,\cite{Degrassi:2012ry}. 
In order to carry out the comparison, one must rewrite the inequality of \cite{Degrassi:2012ry} under a form directly comparable with 
eq.(\ref{eq-boundmt}), namely:
\begin{equation}
m_t^{ [14] %\cite{Degrassi:2012ry}
} [{\rm GeV}] \le 171.8 + 0.5\, (m_H[{\rm GeV}]-126)+ 0.61\, \left( \frac{\alpha_3(m_Z) - 0.1196}{0.0017} \right)
 \pm 0.5_{th} \,.
 \label{eq-14}
\end{equation}
So, our results eq.\,(\ref{eq-boundmt}) and the one obtained in ref.\,\cite{Degrassi:2012ry} are perfectly compatible, as
the central value of eq.\,(\ref{eq-14}) essentially overlaps with the lower value of eq.\,(\ref{eq-boundmt}). 
The theoretical error in eq.\,(\ref{eq-boundmt}) is however smaller than the theoretical error of ref.\,\cite{Degrassi:2012ry}: 
this is mainly due to the fact that in the present analysis we used directly the running top mass,
thus avoiding to introduce the theoretical error due to the matching scale of the top Yukawa coupling (see table 1 
of ref.\,\cite{Degrassi:2012ry}).
The left plot of fig.\,\ref{fig-compIS} shows the comparison between eq.\,(\ref{eq-boundmt}) and eq.\,(\ref{eq-14}) 
in the determination of the transition line between stability and metastability in the $[m_H,m_t]$ plane;
we choose $\alpha_3(m_Z) = 0.1196$ for definiteness, so that the thickness of the lines represents just 
the theoretical error. According to eq.\,(\ref{eq-boundmt}) the thickness of the line is $0.2$ GeV, while
according to eq.\,(\ref{eq-14}) it is $1$ GeV (as can also be checked by inspecting table 1 of \cite{Degrassi:2012ry}).

Clearly all these considerations do not justify the different conclusions of the two papers and rather
show that the different conclusions have to come from the different ranges used for the three most relevant
parameters: $m_t$, $m_H$ and $\alpha_3(m_Z)$. 
In ref.\,\cite{Degrassi:2012ry} it is assumed that $m_t=(173.1 \pm 0.7)$ GeV and $m_H=(125\pm 1)$ GeV; 
these errors are further combined in quadrature and the $1\sigma$ and $2\sigma$ (brown) disks in the right 
plot of fig.\,\ref{fig-compIS} are obtained. 
These disks have to be confronted with our (green) rectangular region,
obtained by using $m_t=(173.3 \pm 2.8)$ GeV, as suggested in ref.\,\cite{Alekhin:2012py}, and $m_H=(125.65 \pm 0.85)$ GeV,
as suggested combining the ATLAS\,\cite{:2012gk} and CMS\,\cite{:2012gu} ranges at $2\sigma$. We have a rectangular
region since we think that in this kind of analysis it is not really justified to combine in quadrature the errors on $m_t$ and $m_H$,
thus enhancing the exclusion of the interesting low-$m_t$ and high-$m_H$ values. 
A small value of $\alpha_3(m_Z)$ also goes in such direction, since it lowers the transition line towards smaller values of $m_t$.
In ref.\,\cite{Degrassi:2012ry} it is assumed that $\alpha_3(m_Z)=0.1184\pm0.0007$: the corresponding
transition line is displayed in the right plot of fig.\,\ref{fig-compIS}, using for consistency eq.\,(\ref{eq-14}).
The line perfectly reproduces the results of fig.\,5 of \cite{Degrassi:2012ry}; its thickness accounts for both the theoretical 
error and the experimental error due to the variation of $\alpha_3(m_Z)$, as derived in \cite{Degrassi:2012ry}. (We cannot
display three separate lines as done in fig.\,\ref{fig-resI} since in eq.\,(\ref{eq-14}) the theoretical error and the one associated to the
variation of $\alpha_3(m_Z)$ are comparable). As the transition line marginally overlaps with the $2\sigma$ disk, the 
authors of ref.\,\cite{Degrassi:2012ry} concluded that stability is disfavored.
A very different conclusion would be derived by considering instead the broad overlap with the rectangle.
This is the main reason of the different conclusions. A small effect is also played by the different values used for $\alpha_3(m_Z)$.  
The range of $\alpha_3(m_Z)$ used in ref.\,\cite{Degrassi:2012ry} has a very small error and has already been questioned 
in ref.\,\cite{Alekhin:2012py} (see bottom of pag 8).
In the present analysis we rather use $\alpha_3(m_Z) = 0.1196 \pm 0.0017$ \cite{PDG}, whose central value and experimental
error are bigger than those used in ref.\,\cite{Degrassi:2012ry}. As an effect, the ensemble of the three (red) lines in fig.\,\ref{fig-resI} 
forms a band slightly wider and higher than the (blue) band in fig.\,\ref{fig-compIS} depicting the results of \cite{Degrassi:2012ry}.

Summarizing, upon comparison of our results in fig.\,\ref{fig-resI} with the results of ref.\,\cite{Degrassi:2012ry} reproduced in the 
right plot of fig.\,\ref{fig-compIS}, one can conclude that the difference in the physical interpretation of the results is mainly due to the fact that
ref.\,\cite{Degrassi:2012ry} adopts a too small experimental error for $m_t$, as already pointed out in ref.\,\cite{Alekhin:2012py}.

%%%%%%%%%%%%%%%%%%%%%%%%%%%%%%%%%%%%%%%%%%

\section{Shallow false minimum}
\label{sec-3}

It is interesting to study in some detail the boundary conditions  which must be satisfied in order to have a very shallow false minimum 
just below the Planck scale, since it could be relevant for inflation \cite{Masina:2011un,Masina:2011aa,Masina:2012yd}.

\begin{figure}[h!]
 \begin{center}
 \vskip .3 cm
 \includegraphics[width=7.5cm]{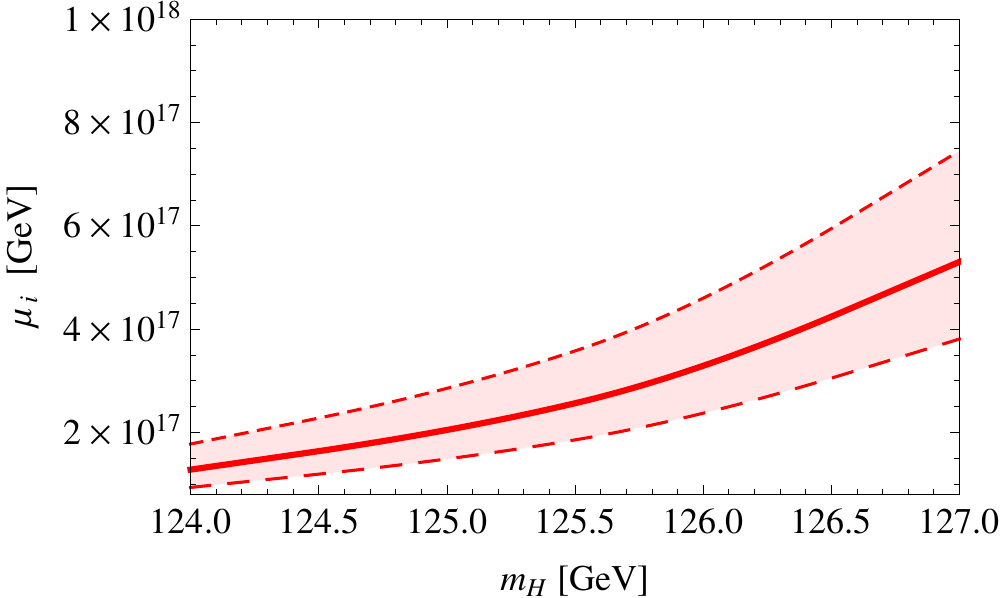} \,\,\,\,\, \,\,\includegraphics[width=7.5cm]{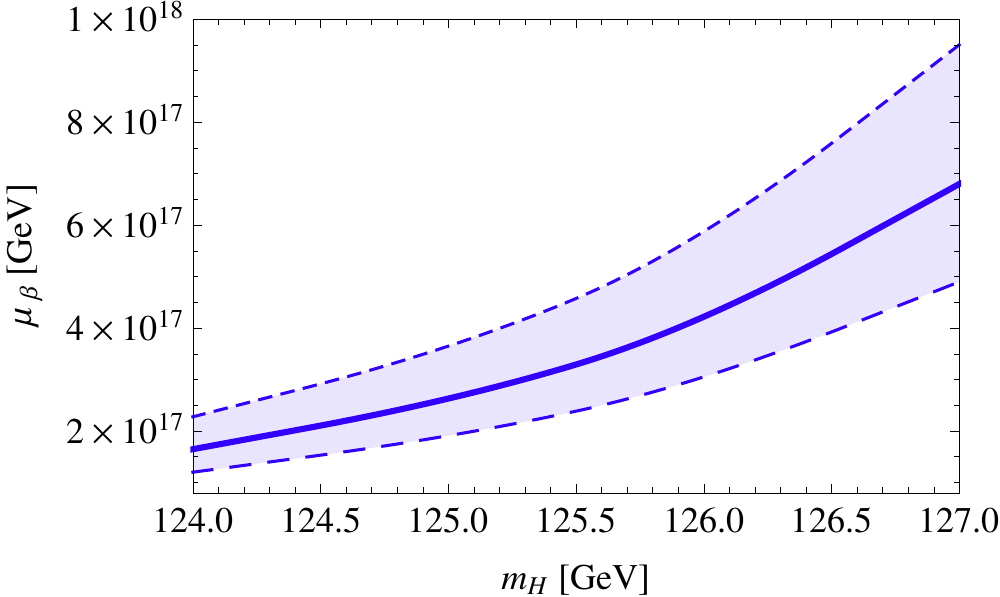}
\end{center}
\caption{\baselineskip=15 pt
Values of $\mu_i$ (left) and $\mu_\beta$ (right) as a function of $m_H$.  For the solid lines, 
the input parameters are fixed at their central values and the matching of $\lambda$ is done at $\mu=m_H$.
The shaded regions shows the uncertainty induced by  the experimental error of $\alpha_3(m_Z)=0.1196 \pm 0.0017$: 
the short and long dashed curves refer to the lower and upper value at $1\sigma$, respectively. 
 }
\label{fig-mu}
\vskip 1 cm
\end{figure}

To study this particular configuration, we denote with $\mu_i$ the renormalization scale where the Higgs potential has an inflection point; 
%, namely $d V^{1/4}/d\mu (\mu_1) =0 $. 
we also recall that $\mu_\beta$ has been defined to be the scale where $\lambda(\mu_\beta)=0$ and $\beta_\lambda(\mu_\beta)=0$ 
are simultaneously fulfilled.
Both $\mu_i$ and $\mu_\beta$ increase\footnote{Notice that $\mu_i$ is slightly smaller
than $\mu_\beta$. This can be easily understood, since
% $ dV^{1/4}/d\mu= 1/4  V^{-3/4}  dV/d\mu$,
%$dV/d\mu= \frac{1}{24} [ 4 \lambda(\mu)  +   d\lambda(\mu)/d\mu  \mu] \mu^3 $.
the condition for having an inflection point at $\mu_i$ reads
$  \beta_\lambda(\mu_i) =- 4 \lambda(\mu_i) <0$, which implies $\mu_i<\mu_\beta$.}
 with $m_H$, as shown in fig.\,\ref{fig-mu}, where the shaded region accounts for the experimental range of $\alpha_3(m_Z)$.

\begin{figure}[h!]
 \begin{center}
 \vskip .3 cm
 \includegraphics[width=7.5cm]{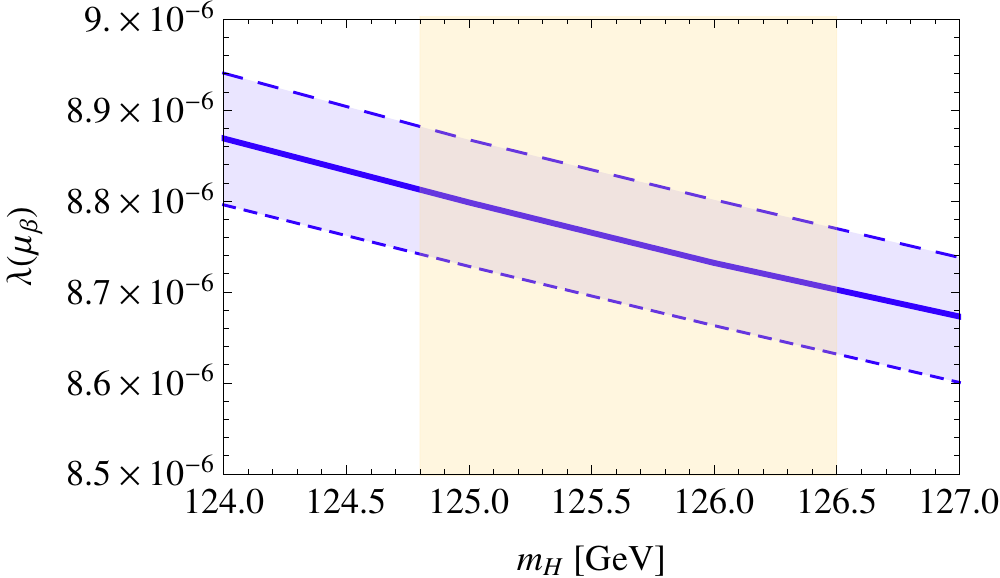}  \,\,\,\,\, \includegraphics[width=7.9cm]{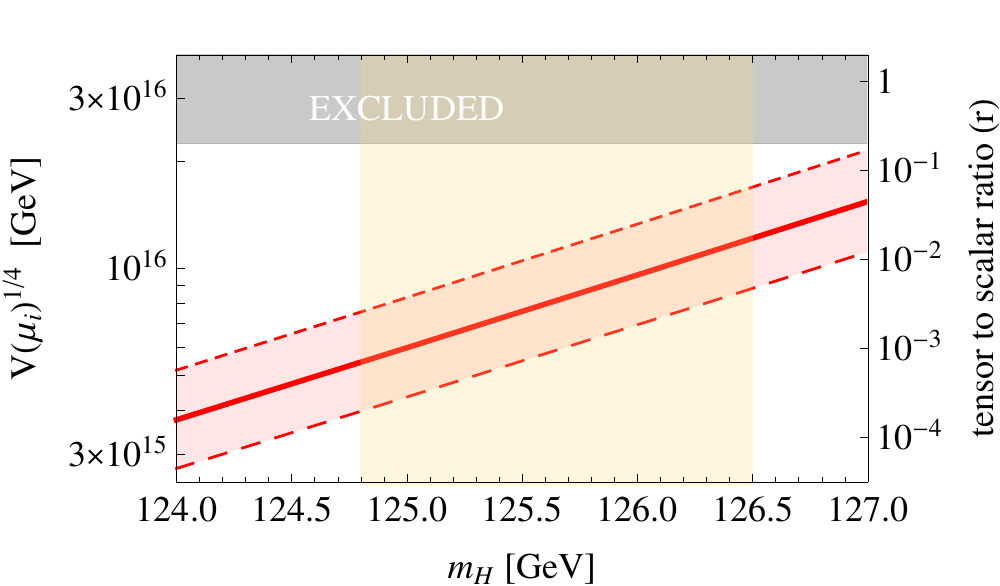} 
\end{center}
\caption{\baselineskip=15 pt
Left: the value of $\lambda(\mu_\beta)$ as a function of $m_H$.
Right: the Higgs potential at $\mu_i$ and the associated prediction for $r$ as a function of $m_H$. 
%The thickness of the lines show the uncertainty induced by  the experimental error of $\alpha_3(m_Z)=0.1196 \pm 0.0017$: 
The short and long dashed curves refer to the $1\sigma$ lower and upper values of $\alpha_3(m_Z)$, respectively.  
The shaded (yellow) vertical region marks the preferred range of $m_H$ at $2\sigma$ \cite{:2012gk,:2012gu}.
The upper region in the right plot is excluded because $r\lesssim 0.2$ \cite{Komatsu:2010fb}.}
\label{fig-lamub}
\vskip .5 cm
\end{figure}
 
 It is interesting that, for the whole experimental range of $m_H$, a shallow false minimum is obtained 
 only if the following boundary condition holds:
 \beq 
 \lambda(\mu_\beta)  \simeq (8.75\pm0.15) \times 10^{-6} \,\,\,\,. \,\, 
 \label{eq-lamub}
 \eeq  
%One could speculate that such a special value could be dictated by an eventual merging with quantum gravity
One could speculate that such value could be originated by some quantum gravity effect
\,\cite{Holthausen:2011aa, Shaposhnikov:2009pv}. 
In the left plot of fig.\,\ref{fig-lamub} we show that $ \lambda(\mu_\beta)$ has a mild dependence on $m_H$; in the right plot we show instead
the value of the Higgs potential at the inflection point, which turns out to be of ${\cal O}(10^{16})$ GeV.  
As before, the shaded regions account for the experimental range of $\alpha_3(m_Z)$.

As pointed out in \cite{Masina:2011un}, a way of testing the hypothesis that inflation occurred when the Higgs field was trapped
into a shallow false vacuum below the Planck scale is to look at the tensor-to-scalar ratio $r$ of cosmological perturbations. 
The amplitude of  density fluctuations in the observed Universe as seen by the CMB and Large-Scale structure data  
is parametrized by the power spectrum in $k$-space,
$P_s(k)=\Delta_R^2 \left( k/k_0 \right)^{n_S-1} $,
where $\Delta_R^2$ is the amplitude at some pivot point $k_0$,
whose best-fit value is $\Delta_R^2= (2.43 \pm 0.11)\times  10^{-9}$  at $k_0=0.002 \,{\rm Mpc}^{-1}$ \cite{Komatsu:2010fb}.
In models where inflation happened while the Higgs was trapped in the shallow minimum \cite{Masina:2011aa,Masina:2012yd}, 
the
Higgs potential at the inflection point and the amount of gravity waves that can be produced - parametrized via 
the tensor-to-scalar ratio $r$ - are linked via a simple relation:
\beq
\Delta_R^2 =  \frac{2}{3 \pi^2}  \frac{1}{r} \frac{ V(\mu_i) }{ M^4  }  \,\,,
\label{eq-r}
\eeq
where $M$ is the reduced Planck scale.
Such prediction for $r$ is reported in the right plot of fig. \ref{fig-lamub}.
Notice that, for these models, only if $m_H$ is in its upper allowed range and $\alpha_3(m_Z)$ is quite low, 
there are chances for the Planck satellite mission \cite{Planck:2006aa} to measure $r$.
However, 
the forthcoming experiment EPIC \cite{Bock:2009xw} should be able to test $r$ down to $10^{-2}$, 
while COrE \cite{Bouchet:2011ck}  down to about $10^{-3}$.

%%%%%%%%%%%%%%%%%%%%%%%%%%%%%%%%%%%%%%%%

\section{Constraints on the Seesaw Mechanism}
\label{sec-4}

We now consider the effect of including neutrino masses via a type I seesaw. 
This issue has been already considered in a series of 
papers \cite{Casas:1999cd,Gogoladze:2008ak,EliasMiro:2011aa,Rodejohann:2012px,Chakrabortty:2012np}.
 
Although the precise amount of the effect is quite model dependent,
here we obtain a conservative estimate of the effect by considering only one right handed neutrino with mass $M_\nu$,
associated to a light Majorana neutrino with mass $m_\nu = 0.06 $ eV,  the scale
of the atmospheric oscillations. This is supported by the following argument.

It is well known that the $\beta$-function of the Higgs quartic coupling is affected only if $h_\nu(\mu)$, 
the Yukawa coupling of the Dirac mass term (defined only for $\mu \ge M_\nu$), is large enough.
As the top Yukawa coupling, also the neutrino Yukawa coupling induces a suppression of the Higgs quartic coupling at high energy. 
By increasing $M_\nu$ and $m_\nu$, the neutrino Yukawa coupling at the threshold scale $M_\nu$ also increases:
\beq
h_\nu(M_\nu) =2  \sqrt{   \frac{ m_\nu(M_\nu)  \, M_\nu}    {v^2}}\,\,.
\eeq
This justifies that the fact that we equate $m_\nu$ to the  the atmospheric mass scale, about $0.06$ eV, which is the lowest possible value
for the heaviest among the three light neutrinos. 
In addition, two other Majorana neutrinos with masses lighter than $m_\nu$ can be accommodated via the seesaw 
but, if their right-handed neutrinos are lighter than $M_\nu$, the associated Dirac Yukawa couplings
are naturally expected to be smaller, and their effect on $\lambda(\mu)$ negligible.

 In Appendix B we provide the additional terms (with respect to the pure SM) for the relevant $\beta$-functions,
 above and below the scale $M_\nu$.

\begin{figure}[t!]
 \begin{center}
 \vskip .3 cm
\includegraphics[width=9.5cm]{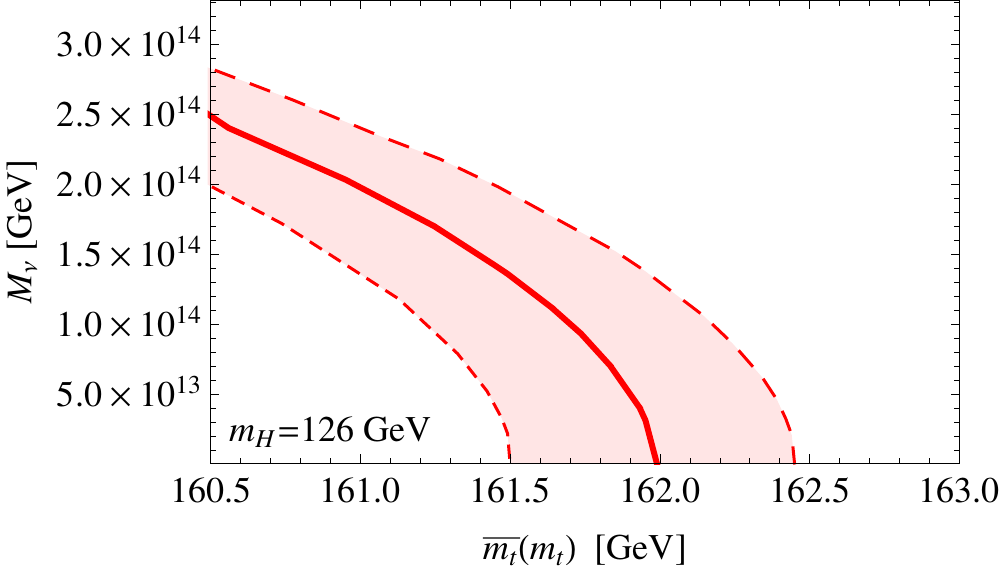}
 \end{center}
\caption{\baselineskip=15 pt
Upper bound on $M_\nu$ as a function of the running top mass, 
following from the requirement that the electroweak vacuum is not destabilized because
of the inclusion of the seesaw, for $m_H=126$ GeV. The shaded region is obtained by varying $\alpha_3(m_Z)$ in its $1\sigma$ range. 
}
\label{fig-M}
\vskip .5 cm
\end{figure}

Since the effect of $h_\nu$ is a suppression of $\lambda$, a SM configuration with a stable electroweak vacuum 
could be rendered metastable because of the addition of the seesaw interactions.
For a fixed value of $m_H$, and in the range of the top mass values allowing the electroweak vacuum to be the global one,
one can find the upper bound on $M_\nu$ following from the requirement that the electroweak vacuum 
remains the global one even after the inclusion of the seesaw interactions. Clearly such upper limit cannot be derived
in the range of the top mass values for which the electroweak vacuum is already metastable.
As shown in fig.\,\ref{fig-M} for $m_H=126$ GeV (but similar upper bounds are obtained in the whole experimental
range of $m_H$), such upper bound strongly depends  on the top mass\footnote{This dependence was not considered 
in the previous literature.} and is affected by an uncertainty which is mainly due to $\alpha_3(m_Z)$ (shaded region). 
The smaller the top mass is, the more the configuration of the Higgs potential is stable and the less stringent 
is the  $M_\nu$ upper bound ensuring that the electroweak vacuum remains the global one and does not become metastable, 
$M_\nu \lesssim 3 \times 10^{14}$ GeV.  But increasing the top mass, the electroweak vacuum becomes less stable and the 
upper bound on $M_\nu$ becomes accordingly more and more stringent. Increasing further the top mass the electroweak vacuum 
becomes metastable even without seesaw interactions, so that no meaningful bound can be derived.

The upper bound on $M_\nu$ following from the requirement of electroweak vacuum stability has to be taken {\it cum grano
salis}, in the sense that it is not a physically robust bound, but just a bound that should be respected in the case one has
a model in which the Higgs potential has to remain stable for reason.

\begin{figure}[t!]
 \begin{center}
 \vskip .3 cm
\includegraphics[width=11cm]{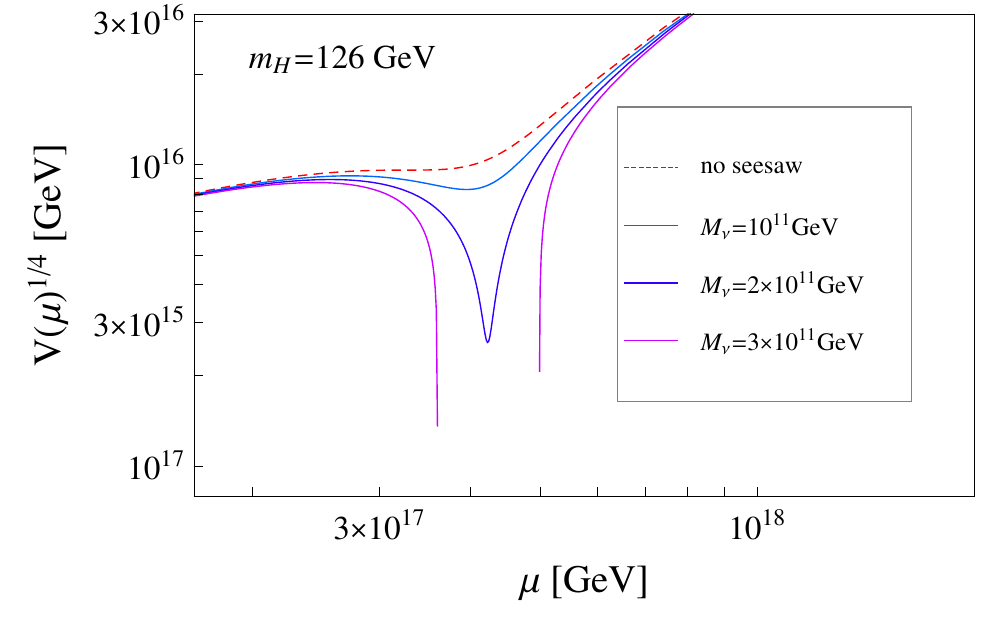} 
 \end{center}
\caption{\baselineskip=15 pt
The dashed curve represents the Higgs potential as  function of the renormalization scale, for $m_H=126$ GeV, 
$\alpha_3(m_Z)=0.1196$ and $m_t = 171.56$ GeV (the value of the top mass leading to an inflection point configuration 
in the SM case once fixed the former two parameters). The solid lower curves display the effect of adding
the seesaw, with three increasing values of $M_\nu$ from top to bottom.}
\label{fig-nu}
\vskip .5 cm
\end{figure}

Let consider in particular the upper bound on $M_\nu$ needed to avoid destabilization of an inflection point  configuration,
as the one depicted via the dashed line in fig.\,\ref{fig-nu}. 
Notice that an inflection point becomes a not so shallow local second minimum if $M_\nu \sim   10^{11}$ GeV and
that electroweak vacuum destabilization is avoided only if the condition $M_\nu \lesssim 2 \times 10^{11}$ GeV
is satisfied. The latter bound might be relevant for  models of inflation based on the SM shallow false minimum 
\cite{Masina:2011un,Masina:2011aa, Masina:2012yd}; note however that it is well compatible with the thermal leptogenesis mechanism
to explain matter-antimatter asymmetry, for which the lower bound on the lightest Majorana neutrino is about $5\times 10^{8}$ GeV
\cite{Davidson:2002qv}.

Clearly, the neutrino Yukawa coupling $y_\nu$ is not the only additional term beyond the SM capable of modifying 
the running of $\lambda$ at high energy. Always in the context of type I seesaw, in the case that the vacuum expectation value
of a singlet scalar field $S$ (violating the lepton number by two units) is actually at the origin of the right-handed Majorana neutrino mass, 
the $S$ couplings induce an enhancement of $\lambda$, thus helping the stability of the electroweak vacuum \cite{EliasMiro:2012ay}.
Such effect is indeed generically expected when adding to the SM  a singlet field $S$ \cite{Lebedev:2012zw, EliasMiro:2012ay}.

%%%%%%%%%%%%%%%%%%%%%%%%%%%%

\section{Conclusions}
\label{sec-concl}

The recent discovery of a particle consistent with the SM Higgs boson \cite{:2012gk,:2012gu} 
provides a strong motivation to pursue \cite{Holthausen:2011aa,Masina:2011aa,EliasMiro:2011aa,Xing:2011aa,
Masina:2011un,Bezrukov:2012sa,Degrassi:2012ry,Alekhin:2012py} 
the old project \cite{1group, 2group,3group} 
of investigating the behavior of the SM Higgs potential at very high energies. 
In particular, one would understand whether the SM electroweak vacuum
is a global minimum up to the Planck scale, namely whether we live in a stable vacuum assuming a desert 
(or assuming that new interactions do not modify the running of $\lambda$ with respect to the SM case).
In particular, a stable configuration which deserves a special interest is a shallow 
false minimum below the Planck scale: the Higgs field could have been primordially trapped there, 
leading to a stage of inflation\,\cite{Masina:2011aa,Masina:2011un,Masina:2012yd}.
 Stability below the Planck scale is required also in Higgs inflation models with 
 non minimal gravitational couplings \,\cite{Bezrukov:2007ep,Bezrukov:2012sa}. 

In our analysis, we adopted the recently derived tools for a NNLO renormalization procedure\,\cite{Mihaila:2012fm,Chetyrkin:2012rz,Bezrukov:2012sa,Degrassi:2012ry}.  At difference of previous analyses, 
we considered as free parameter the running top mass rather than the (Tevatron) top pole mass, as suggested in ref.\,\cite{Alekhin:2012py}.

Given the present range of the running top mass and of the Higgs mass, we found that electroweak vacuum stability is at present allowed,    
as shown in figs.\,\ref{fig-line} and \ref{fig-resI}. 
To further test stability, a more precise measurement of the top mass would be crucial. 
As apparent from the stability condition of eq.\,(\ref{eq-boundmt}), 
in case that LHC will not exclude values of the running top mass below $163$ GeV (or equivalently
 values of the pole top mass below $173$ GeV), 
an electron-positron collider would probably be needed to discriminate between stability and metastability.

We also determined the high scale boundary conditions allowing for a shallow false minimum slightly below the Planck scale, 
$\lambda(\mu_\beta)  \sim 10^{-5}$ ($\mu_\beta$ is the renormalization scale were the $\beta$-function of
 the Higgs quartic coupling vanishes), and discussed the prospects for the cosmological tests of such configuration.
Finally, a conservative upper bound on type I seesaw right-handed neutrino masses, following from the requirement of electroweak 
vacuum stability, was derived, analyzing in particular its dependence on the top mass.

The present analysis does not consider the effect of the gravitational couplings because it is far from
clear how the quantum effects of the latter would impact the study at very high energies. 
Other sources of uncertainty could also come from the treatment of the effective potential itself, 
such as the fine-tuning required for both the cosmological constant and the Higgs mass.

%%%%%%%%%%%%%%%%%%%%%%%%%

\vskip 1 cm

\section*{Acknowledgments}We would like to thank G. Isidori, G. Moore, A. Notari and A. Strumia for useful discussions.

%%%%%%%%%%%%%%%%%%%%%%%%%%%%%%%%%%%%%%
\appendix
\section{Formul\ae \, for the RG running at NNLO}
\label{app-a}

%%%%%%%%%%%%
\subsection{The $\beta$-functions}

Here we provide the expressions for the $\beta$-functions up to 3-loops, see eq. (\ref{eq-RGE}).

At 1-loop they are given by:
\begin{eqnarray}
\nonumber
\beta_\lambda^{(1)}&=&\frac{27}{4} g(t)^4+\frac{9}{2} g'(t)^2 g(t)^2-9 \lambda(t) g(t)^2+\frac{9}{4} g'(t)^4-36 h_t(t)^4+4 \lambda(t)^2-3 g'(t)^2 \lambda(t) \\ 
   && +12 h_t(t)^2 \lambda(t)\  \nonumber, \\  
\nonumber
\beta^{(1)}_{h_t}&=& \frac92 h_t(t)^3-\frac{9}{4} g(t)^2 h_t(t)-8 g_3(t)^2 h_t(t)-\frac{17}{12} g'(t)^2 h_t(t)\ , \\ 
\nonumber \beta_g^{(1)}&=&-\frac{19}{6}   g(t)^3\ ,\\  
\nonumber \beta_{g'}^{(1)}&=&\frac{41}{6}   g'(t)^3\ ,\\ 
\nonumber \beta_{g_3}^{(1)}&=&-7  g_3(t)^3\ .
\end{eqnarray}

At 2-loop they are:
\begin{eqnarray}
\nonumber 
\beta_\lambda^{(2)}&=&80g_3(t)^2 h_t(t)^2 \lambda (t)-192g_3(t)^2 h_t(t)^4+ 
   \frac{915}{8} g(t)^6-\frac{289}{8} g'(t)^2 g(t)^4-\frac{27}{2} h_t(t)^2 g(t)^4\\  \nonumber &&  
   -\frac{73}{8} \lambda (t) g(t)^4-\frac{559}{8} g'(t)^4 g(t)^2+63 g'(t)^2 h_t(t)^2 g(t)^2+\frac{39}{4} g'(t)^2 \lambda (t)
   g(t)^2-3 h_t(t)^4 \lambda (t)\\ &&  \nonumber +\frac{45}{2} h_t(t)^2 \lambda (t) g(t)^2
   -\frac{379}{8} g'(t)^6+180 h_t(t)^6-16 g'(t)^2 h_t(t)^4-\frac{26}{3}\lambda (t)^3
   -\frac{57}{2} g'(t)^4 h_t(t)^2\\ &&  \nonumber -24 h_t(t)^2 \lambda (t)^2   +6 \left(3 g(t)^2+g'(t)^2\right) \lambda
   (t)^2+\frac{629}{24} g'(t)^4 \lambda (t)+\frac{85}{6} g'(t)^2 h_t(t)^2 \lambda (t)
   \ , 
   \end{eqnarray}
   \begin{eqnarray}
   \nonumber
\beta_{h_t}^{(2)}&=&h_t(t) \left[-108g_3(t)^4+9 g(t)^2g_3(t)^2+\frac{19}{9} g'(t)^2g_3(t)^2+36 h_t(t)^2
  g_3(t)^2-\frac{3}{4} g'(t)^2 g(t)^2\right. \\ && \nonumber
  \left. -\frac{23}{4} g(t)^4+\frac{1187}{216}g'(t)^4-12 h_t(t)^4+\frac{\lambda (t)^2}{6}+h_t(t)^2 \left(\frac{225}{16} g(t)^2+\frac{131}{16} g'(t)^2-2
   \lambda (t)\right)\right] \ ,\\ \nonumber
 \end{eqnarray}
\begin{eqnarray}
\beta_{g}^{(2)}&=&12 g_3(t)^2 g(t)^3+   \left(\frac{35}{6} g(t)^2+\frac32 g'(t)^2-\frac32 h_t(t)^2\right) g(t)^3\ ,\nonumber \\ \nonumber
\beta_{g'}^{(2)}&=&\frac{44}{3} g_3(t)^2 g'(t)^3+   \left(\frac92 g(t)^2+\frac{199}{18} g'(t)^2-\frac{17}{6} h_t(t)^2\right) g'(t)^3\ ,\\ \nonumber
\beta_{g_3}^{(2)}&=&g_3(t)^3 \left(\frac92 g(t)^2-26g_3(t)^2+\frac{11}{6} g'(t)^2-2 h_t(t)^2\right)\ .
 \end{eqnarray}

The leading terms in the 3-loop $\beta$-functions of $\lambda$ and $h_t$ are \cite{Chetyrkin:2012rz}:
\begin{eqnarray}
\nonumber
   \beta_{\lambda}^{(3)}&=&12 \left[  (-\frac{266}{3} + 32 \zeta_3) g_3(t)^4 h_t(t)^4 + (-38 + 240 \zeta_3) g_3(t)^2 h_t(t)^6 
    - (\frac{1599}{8} + 36 \zeta_3) h_t(t)^8 \right. \nonumber \\
   &&+ \frac{1}{6} (\frac{1244}{3} - 48 \zeta_3) g_3(t)^4 h_t(t)^2 \lambda(t) + 
   \frac{1}{6} (895 - 1296 \zeta_3) g_3(t)^2 h_t(t)^4 \lambda(t) + \nonumber \\
   && \frac{1}{6} (\frac{117}{8} - 198 \zeta_3) h_t(t)^6 \lambda(t) + 
   \frac{1}{36} (-1224 + 1152 \zeta_3) g_3(t)^2 h_t(t)^2 \lambda(t)^2 + \nonumber \\
  && \frac{1}{36} (\frac{1719}{2} + 756 \zeta_3) h_t(t)^4 \lambda(t)^2 + 
 \left.  \frac{ 97}{24} h_t(t)^2 \lambda(t)^3 + \frac{1}{1296} (3588 + 2016 \zeta_3) \lambda(t)^4 \right]  \, ,\nonumber
\\
\nonumber
   \beta_{h_t}^{(3)}&=&    %384 g_3(t)^6 \left(-\frac{2083}{576}+\frac53\zeta_3\right)\ ,\\ 
   2 \left[ (-\frac{2083}{3} + 320 \zeta_3) g_3(t)^6 + (\frac{3827}{12} - 114 \zeta_3) g_3(t)^4 h_t(t)^2 - 
   \frac{157}{2} g_3(t)^2 h_t(t)^4  \right. \\ 
   && + (\frac{339}{16} + \frac{27}{4} \zeta_3) h(t)^6 +  \frac{4}{3} g_3(t)^2 h_t(t)^2 \lambda(t) + \frac{33}{2} h_t(t)^4 \lambda(t) + 
 \left.   \frac{5}{96} h_t(t)^2 \lambda(t)^2 -\frac{1}{12} \lambda(t)^3 \right] \, ,\nonumber
\end{eqnarray}
where $\zeta_3=1.20206...$ is the Riemann zeta function.
   
The complete 3-loop $\beta$-functions for the gauge couplings are \cite{Mihaila:2012fm}:
 \begin{eqnarray}
 \beta_{g}^{(3)}&=& \frac{324953}{1728} g(t)^7 + 39 g(t)^5 g_3(t)^2 + 81 g(t)^3 g_3(t)^4 + 
 \frac{291}{32} g(t)^5 g'(t)^2 - \frac{1}{3} g(t)^3 g_3(t)^2 g'(t)^2  \nonumber \\
 && - \frac{ 5597}{576} g(t)^3 g'(t)^4 - \frac{729}{32} g(t)^5 h_t(t)^2 - 
 7 g(t)^3 g_3(t)^2 h_t(t)^2 - \frac{593}{96} g(t)^3 g'(t)^2 h_t(t)^2 \nonumber \\
 &&+ \frac{ 147}{16} g(t)^3 h_t(t)^4 \, ,  \nonumber \\
 \beta_{g'}^{(3)}&=& \frac{1315}{64} g(t)^4 g'(t)^3 - g(t)^2 g_3(t)^2 g'(t)^3 +  99 g_3(t)^4 g'(t0^3 + \frac{205}{96} g(t)^2 g'(t)^5 - 
\frac{ 137}{27} g_3(t)^2 g'(t)^5 \nonumber \\
&& - \frac{388613}{5184} g'(t)^7 - \frac{ 785}{32} g(t)^2 g'(t)^3 h_t(t)^2 - \frac{29}{3} g_3(t)^2 g'(t)^3 h_t(t)^2 
- \frac{ 2827}{288} g'(t)^5 h_t(t)^2 \nonumber \\
&&+ \frac{315}{16} g'(t)^3 h_t(t)^4 \, , \nonumber
 \\
   \nonumber
\beta_{g_3}^{(3)}&=& \frac{109}{8} g(t)^4 g_3(t)^3 + 21 g(t)^2 g_3(t)^5 + \frac{65}{2} g_3(t)^7 -  \frac{1}{8} g(t)^2 g_3(t)^3 g'(t)^2 
+ \frac{77}{9} g_3(t)^5 g'(t)^2 \, \nonumber \\ 
&&- \frac{ 2615}{216} g_3(t)^3 g'(t)^4 -\frac{ 93}{8} g(t)^2 g_3(t)^3 h_t(t)^2 - 
 40 g_3(t)^5 h_t(t)^2 + \frac{101}{24} g_3(t)^3 g'(t)^2 h_t(t)^2 \, \nonumber \\
 &&+ 15 g_3(t)^3 h_t(t)^4 \, .\nonumber
\end{eqnarray}

%%%%%%%%%%%%%%
\subsection{Higgs quartic coupling matching}

According to Sirlin and Zucchini~\cite{SZ}, the 1-loop matching is given by 
\begin{equation}
\delta^{(1)}_H (\mu)=\frac{G_\mu m_Z^2}{8 \sqrt{2} \pi ^2} \left(\xi  f_1(\mu)+f_0(\mu)+\frac{f_{-1}(\mu)}{\xi}\right) \, ,\nonumber
\end{equation}
where
$\xi =\frac{m_H^2}{m_Z^2}$ and, introducing $ c=\frac{m_W}{m_Z}$, 
\begin{eqnarray}
f_1(\mu)&=&\frac{3}{2} \log (\xi )-\log \left(c^2\right)+6 \log \left(\frac{\mu^2}{m_H^2}\right)-\frac{1}{2}
   Z\left[\frac{1}{\xi }\right]-Z\left[\frac{c^2}{\xi }\right]+\frac{9}{2} \left(\frac{25}{9}-\frac{\pi }{\sqrt{3}}\right)\,, \nonumber\\
f_0(\mu)&=&\frac{3  c^2}{s^2}\log \left(c^2\right)+12 \log c^2\left(c^2\right) +\frac{3 \xi   c^2}{\xi -c^2}\log \left(\frac{\xi
   }{c^2}\right)+4c^2\,Z\left[\frac{c^2}{\xi }\right] -\frac{15}{2} \left(2 c^2+1\right) \nonumber\\
   \nonumber && -6 \left(2 c^2-\frac{2m_t^2}{m_Z^2}+1\right) \log \left(\frac{\mu^2}{m_Z^2}\right)
   -\frac{3m_t^2}{m_Z^2} \left(4 \log \left(\frac{m_t^2}{m_Z^2}\right)
   +2\,Z\left[\frac{m_t^2}{m_Z^2 \xi}\right]-5\right) \nonumber\\
   &&+2\, Z\left[\frac{1}{\xi }\right]\,,\nonumber \\ \nonumber 
f_{-1}(\mu)&=&8 \left(2 c^4+1\right)-12c^4 \log \left(c^2\right) -12c^4\, Z\left[\frac{c^2}{\xi }\right] +6 \left(2
   c^4-\frac{4 m_t^4}{m_Z^4}+1\right) \log \left(\frac{\mu^2}{m_Z^2}\right)\nonumber\\
   &&-6\,Z\left[\frac{1}{\xi }\right]+\frac{24 m_t^4 }{m_Z^4}\left(\log \left(\frac{m_t^2}{m_Z^2}\right)+Z\,\left[\frac{m_t^2}{m_Z^2
   \xi }\right]-2\right)\,,\nonumber\\
Z[z]&=&\left\{ \begin{array}{lll}
2 A(z) \,{\rm arctan}\left(\frac{1}{A(z)}\right) &  \textrm{if} & z>\frac{1}{4}\nonumber \\
A(z) \log \left(\frac{A(z)+1}{1-A(z)}\right) &  \textrm{if} & z<\frac{1}{4}\nonumber
\end{array}
\right. \,, \\
A(z)&=&\sqrt{|1-4 z|} \,.\nonumber
\end{eqnarray}

We compute  the QCD and the Yukawa contribution to $\lambda^{(2)}(\mu)$ following the expressions of \cite{Degrassi:2012ry}
(multiplied them by a factor of $6$ to compensate for the different definition of the quartic coupling). 

\vskip 1cm

%%%%%%%%%%%%%%%%%%%%%%%%%%%%%%%%%%%%%%%%%%%

\section{Seesaw contribution to the $\beta$-functions}
\label{app-b}

Below the right handed neutrino mass scale, the running of the effective light Majorana neutrino mass is given by \cite{Babu:1993qv}
\beq
\frac{d m_\nu(t)}{dt} = \kappa  \left(- 3 g_2(t)^2 + 6 h_t(t)^2 + \frac{\lambda(t)}{6} \right) m_\nu(t) \,\,.\nonumber
\eeq

For $\mu > M_\nu$, we have \cite{Pirogov:1998tj}
\beq
\frac{ d h_\nu(t)}{dt} =  \kappa \, h_\nu(t) \left(\frac{5}{4} h_\nu(t)^2 + \frac{3}{2} h_t(t)^2 - \frac{3}{4} g'(t)^2 - \frac{9}{4} g(t)^2\right) \,\,,\nonumber
\eeq
together with
\beq
\delta \beta^{(1)}_\lambda =- 3 h_\nu(t)^4 + 2 \lambda(t) h_\nu(t)^2 \,\,\,,\,\,\, \delta \beta^{(1)}_{h_t}=  \frac{1}{2} h_\nu(t)^2  \,\, .\nonumber
\eeq

%%%%%%%%%%%%%%%%%%%%%%%%%%%%%%%%%
%%%%%%%%%%%%%%%%%%%%%%%%%%%%%%%%%

%\newpage

\end{document}